\documentclass[animals,article,accept,pdftex,moreauthors]{Definitions/mdpi} 
\firstpage{1} 
\pubvolume{1}
\issuenum{1}
\articlenumber{0}
\pubyear{2025}
\copyrightyear{2025}
\externaleditor{Cédric Sueur} 
\datereceived{18 February 2025} 
\daterevised{13 March 2025} 
\dateaccepted{17 March 2025} 
\datepublished{ } 
\hreflink{https://doi.org/} 

\usepackage{changepage}

\Title{Predicting Dairy Calf Body Weight from Depth Images Using Deep Learning (YOLOv8) and Threshold Segmentation with Cross-Validation and Longitudinal Analysis}

\TitleCitation{Predicting Dairy Calf Body Weight from Depth Images Using Deep Learning (YOLOv8) and Threshold Segmentation with Cross-Validation and Longitudinal Analysis}

\Author{Mingsi Liao $^{1}$\orcidA{}, Gota Morota $^{1,2}$\orcidB{}, Ye Bi $^{1}$\orcidC{} and Rebecca R.  Cockrum $^{1,}$*\orcidD{}}

\AuthorNames{Mingsi Liao, Gota Morota, Ye Bi and Rebecca R. Cockrum}

\AuthorCitation{Liao, M.; Morota, G.; Bi, Y.; Cockrum, R.R.}

\address{%
$^{1}$ \quad School of Animal Sciences, Virginia Polytechnic Institute and 
 State University, Blacksburg, 24060 VA, USA; msliao@vt.edu (M.L.); gmorota@g.ecc.u-tokyo.ac.jp (G.M.); yebi@vt.edu (Y.B.)  \\
$^{2}$ \quad Laboratory of Biometry and Bioinformatics, Department of Agricultural and Environmental Biology, Graduate School of Agricultural and Life Sciences, The University of Tokyo, Bunkyo, {Tokyo} 113-8657, 
 Japan}

\corres{Correspondence: rcockrum@vt.edu}

\simplesumm{Weighing dairy calves is essential for monitoring their health and growth, but traditional methods require manual effort, which can be time-consuming for farm workers and stressful for animals. This study explores a new way to estimate calf weight using images and artificial intelligence, allowing for automatic and non-invasive weight prediction. We compared two methods for processing calf images: a simple approach that relies on rule-based image processing and a more advanced deep learning technique. Our results showed that deep learning provided more accurate body measurements, leading to better weight predictions. We tested different models to predict calf growth and found that incorporating a linear mixed model most accurately forecasted weight over time. Our findings suggest that body measurements taken at a young age can help predict future weight, which can assist farmers in making better decisions about feeding and managing calves. This approach can make weight monitoring more efficient and improve animal welfare by reducing handling stress. By integrating artificial intelligence into livestock management, this study provides a valuable tool for improving farm productivity and ensuring the healthy growth of dairy calves.} 

\abstract{Monitoring calf body weight (BW) before weaning is essential for assessing growth, feed efficiency, health, and weaning readiness. However, labor, time, and facility constraints limit BW collection. {Additionally, Holstein calf coat patterns complicate image-based BW estimation, and few studies have explored non-contact measurements taken at early time points for predicting later BW.} The objectives of this study were to (1) develop deep learning-based segmentation models for extracting calf body metrics, (2) compare deep learning segmentation with threshold-based methods, and (3) evaluate BW prediction using single-time-point cross-validation with linear regression (LR) and extreme gradient boosting (XGBoost) and multiple-time-point cross-validation with LR, XGBoost, and a linear mixed model (LMM). Depth images from Holstein (n = 63) and Jersey (n = 5) pre-weaning calves were collected, with 20 Holstein calves being weighed manually. Results showed that You Only Look Once version 8 (YOLOv8) deep learning segmentation (intersection over union = 0.98) outperformed threshold-based methods (0.89). In single-time-point cross-validation, XGBoost achieved the best BW prediction (R$^2$ = 0.91, mean absolute percentage error (MAPE) = 4.37\%), while LMM provided the most accurate longitudinal BW prediction (R$^2$ = 0.99, MAPE = 2.39\%). These findings highlight the potential of deep learning for automated BW prediction, enhancing farm management.\vspace{6pt}}


\begin{document}


\section{Introduction}

\textls[-25]{Body weight (BW) is an important measure for dairy calf management, serving as the foundation for management decisions related to health, feeding, and~overall lifetime productivity~\cite{appleby2001performance,bradski2000opencv}. Traditionally, calf BW has been measured using electronic scales, which can vary in precision and accuracy. Furthermore, unlike other livestock, calf working facilities are limited to individually positioning calves on the scale rather than having access to a chute where weights could be collected more efficiently. The consistent collection of calf weights is time-consuming, labor-intensive, and~stressful for the animals, which may explain why calf weights are rarely collected on dairy farms~\cite{sharpe2023evaluation,song2014body,cantor2020estimating}. Early efforts focused on developing mathematical models that use easy-to-measure physical characteristics, such as body girth and height, to predict BW~\cite{heinrichs1992predicting, jeong1998study}. Although~these methods reduced the need for direct weighing, they still required manual measurements and are subject to variability depending on the skill and consistency of the operators. To~address these challenges, researchers have sought alternative methods to estimate the BW of dairy cattle. Advancements in technology have led to the use of video cameras for the prediction of cattle BW. For~example, a~Canon Mv850i video camera was used to capture two-dimensional (2D) images to predict the BW of beef cattle~\cite{bozkurt2007body}. This approach represented a significant step toward streamlining the measurement process, reducing the need for direct physical interaction with the animals. Subsequently, \citet{stajnko2010non} introduced thermal imaging as another non-invasive technique to estimate the live weight of cattle. Researchers also explored the potential of depth cameras, which offered more precise estimates. Specifically, \citet{gjergji2020deep} demonstrated that depth image analysis could effectively estimate both the BW and body composition of beef cattle, offering a non-invasive and potentially more accurate method for monitoring~livestock. }

Image analysis has enabled the automatic prediction of BW using threshold-based segmentation, a~technique that separates objects from the background by applying a fixed pixel intensity threshold~\cite{kadlec2022automated}. These methods marked a significant advancement in animal management, allowing the more frequent and less invasive monitoring of cattle. Threshold-based approaches are also efficient in image processing and require fewer computational resources, making them suitable for resource-constrained environments. However, threshold-based approaches are limited in their ability to accurately segment images varying in environmental conditions. These shortcomings often lead to inaccurate body metrics and, consequently, less reliable BW predictions. Given these limitations, the~application of deep learning techniques has emerged as a promising alternative. Furthermore, many existing studies focus on adult cattle, with~limited research that addresses the unique challenges of predicting BW in dairy calves, where rapid growth and smaller body sizes pose additional complexities. Furthermore, beef cattle used for previous studies have uniform coloring, whereas the~color patching in Holstein's calves introduces an additional hurdle to using visual imaging to predict BW. This study seeks to fill these gaps by exploring novel methodologies that leverage advanced deep learning and machine learning techniques specifically tailored for Holstein heifer~calves.

Deep learning models, particularly convolutional neural networks (CNNs), have shown superior performance in various image analysis tasks due to their ability to learn complex patterns and features from large datasets~\cite{gjergji2020deep,bi2023depth}. The~YOLO (You Only Look Once) family of models~\cite{redmon2016you} is a highly effective CNN approach known for its excellence in real-time object detection and segmentation tasks. A~key advantage of using YOLOv8 lies in its pre-training process, where the model is first trained on a large and diverse dataset such as COCO and ImageNet. This pre-training allows the model to learn general features that are applicable across different tasks, providing a strong foundation that can be fine-tuned for specific applications. You Only Look Once version 8 comes in different sizes, namely n (nano), s (small), m (medium), l (large), and~x (extra-large). The~smaller versions, n and s, are faster but less accurate, whereas m and l versions offer greater accuracy at the cost of requiring more computational power. While deep learning models excel in object detection and segmentation, structured machine learning models such as extreme gradient boosting (XGBoost) offer powerful capabilities for numerical prediction, including BW~estimation.

Extreme gradient boosting is a powerful and widely used machine learning algorithm. Developed by \citet{chen2016xgboost}, XGBoost builds on the concept of gradient boosting, an~ensemble learning technique that combines the predictions of multiple decision trees to form a strong predictive model. Each tree is sequentially added to the model, as~each new tree corrects the errors made by the previous ones, thereby improving the overall accuracy of the model. One of the key strengths of XGBoost is its regularization techniques, L1 (LASSO) and L2 (Ridge) regularization, which help prevent overfitting, a common issue in machine learning where the model performs well on the training data but poorly on unseen data. Additionally, XGBoost is highly efficient in handling large datasets, missing values, and~various data types, making it a versatile tool in predictive analytics. While XGBoost provides a powerful tool for single-time-point BW prediction, understanding how calf BW evolves over time requires statistical methods designed to model repeated measurements. Longitudinal analysis or repeated measures analysis offers a framework to capture these dynamic changes, ensuring more accurate assessments of growth and~development.

\textls[-25]{Longitudinal analysis is a statistical approach used to study repeated measurements collected over time from the same subjects. This method captures individual growth patterns, developmental changes, and~responses to interventions, making it particularly useful in animal science and biomedical research~\cite{singer2003applied}. In~dairy calves, early-life nutrition and housing conditions influence long-term growth and reproductive performance, with~key factors including milk feeding strategies, housing type, pre-weaning growth rates, and~disease incidence~\cite{curtis2018impact}. Therefore, modeling longitudinal data using appropriate statistical methods is essential for understanding these relationships and making data-driven management decisions. One widely used approach for analyzing longitudinal data is the linear mixed model (LMM) \cite{laird1982random}, first introduced by \citet{henderson1959estimation}. Linear mixed models are particularly useful for handling repeated measures because they account for both fixed effects (e.g., dietary treatment, housing) and random effects (e.g., individual variability among calves). These models allow researchers to properly model within-subject correlations and account for missing data, making them a powerful tool for studying time-dependent biological~processes.}

The objectives of this study were to (1) develop and validate deep learning-based segmentation models (YOLOv8) for extracting calf body metrics from dorsal depth images, (2) compare the performance of the deep learning model with threshold-based image processing techniques in terms of segmentation accuracy, and~(3) evaluate BW prediction accuracy using single-time-point cross-validation with linear regression (LR) and XGBoost, as~well as multiple-time-point cross-validation with an LMM to capture individual growth trajectories over~time.

\section{Materials and~Methods}
\unskip
\subsection{Animal Procedures, Data Collection, and~Image~Acquisition}
\begin{figure}[H]
\includegraphics[width=13cm]{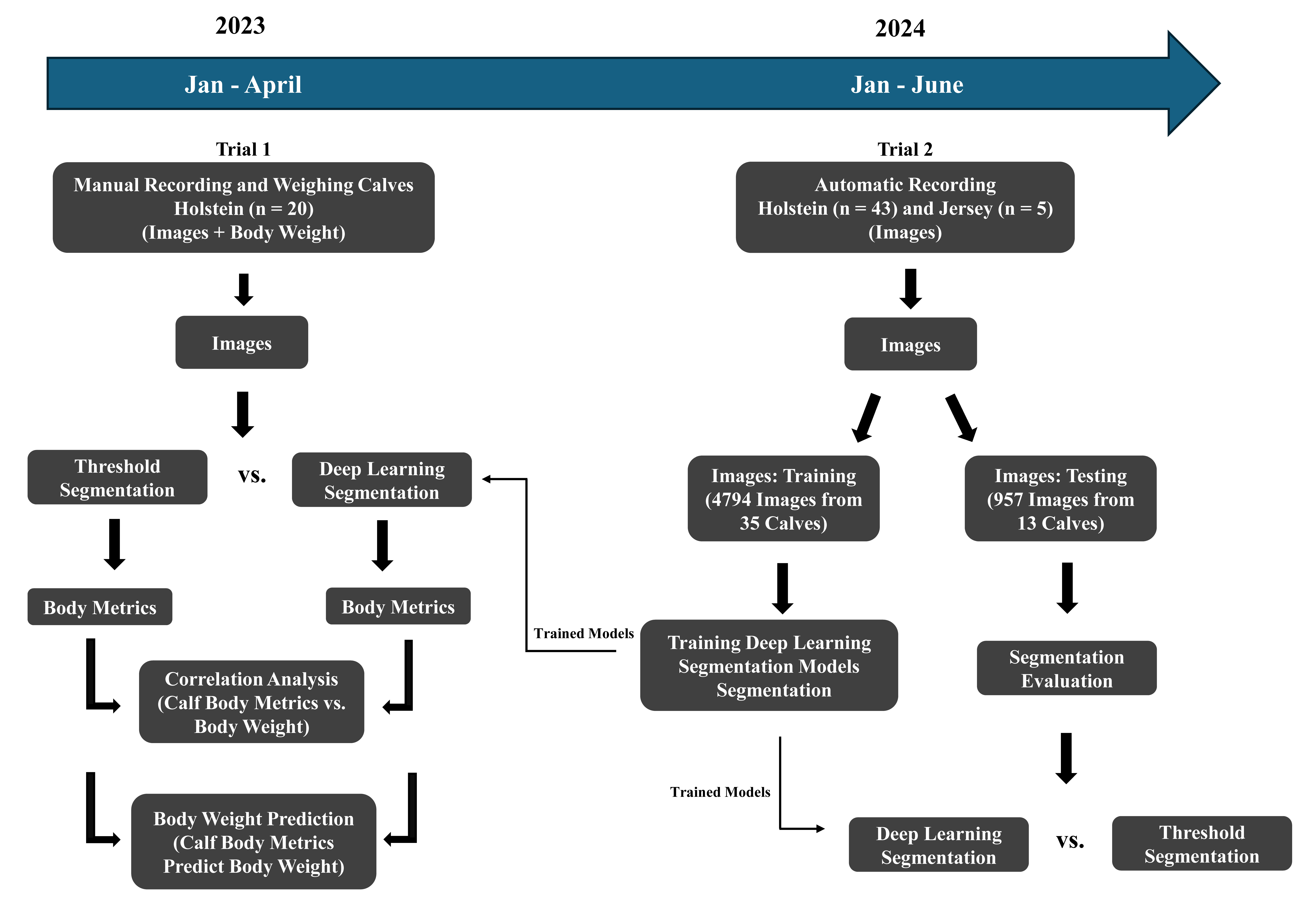}
\caption{{Flowchart of the dairy calf body weight prediction process.}}
\label{fig1}
\end{figure}

The animal procedures were approved by the Virginia Polytechnic Institute and State University Animal Care and Use Committee (Protocol \#22-197). Data were collected on Holstein calves (trial 1; n = 20) for correlation and prediction analysis between calf body metrics and weight, as well as Holstein and Jersey calves (trial 2; n = 48) for developing segmentation models using deep learning ({Figure~\ref{fig1};} Table~\ref{tab1}). There was no overlap between the animals in these two trials. Calves were group-housed in a covered barn at the Virginia Tech Dairy Complex. All calves were fed by an automatic feeder (Forster Technik,  Engen, Germany
).

\begin{figure}[H]
\includegraphics[width=13cm]{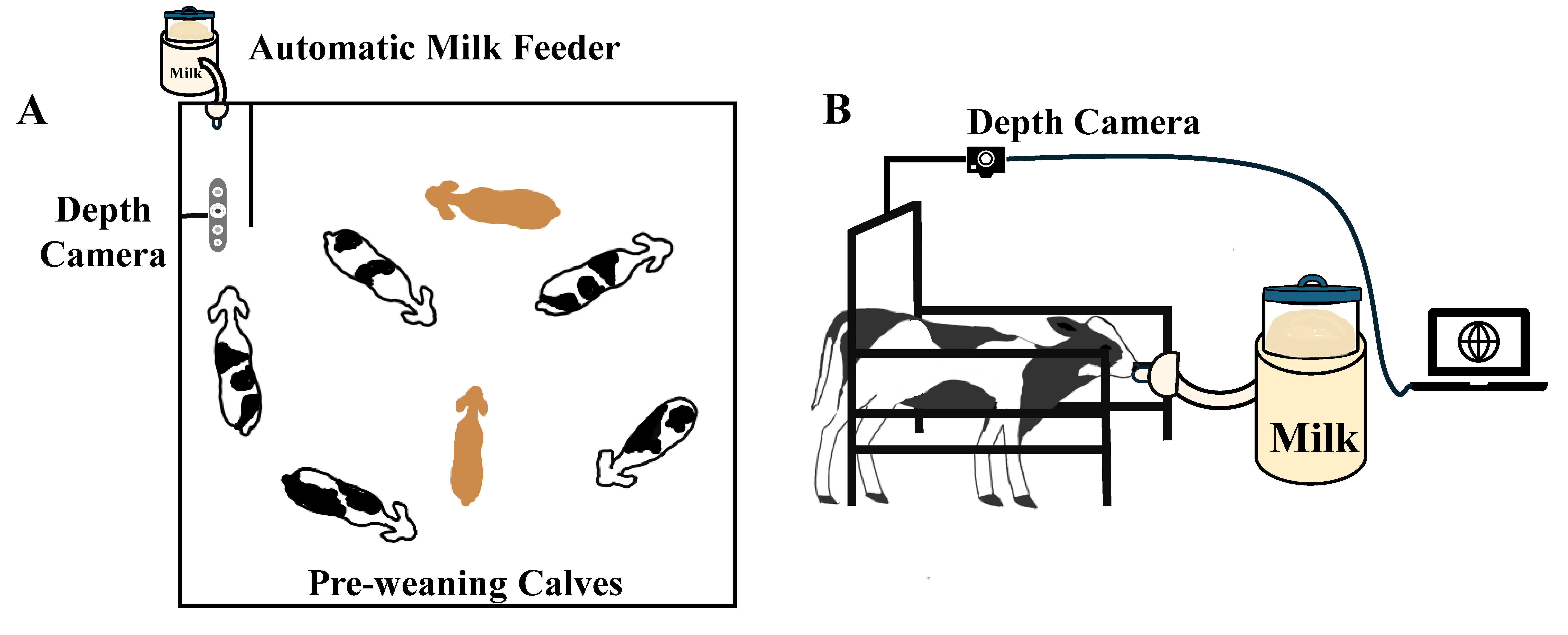}
\caption{{(\textbf{A}) Top-view of the calf pen equipped with an automatic milk feeder and a depth camera. (\textbf{B}) Side-view of a dairy calf during milk feeding, with~a depth camera capturing videos.}}
\label{fig2}
\end{figure}

In trial 1, Holstein calves (n = 20) ranged in age from 21 to 69 days, with~an average BW of 146.73 $\pm$ 26.31 lb (66.54 $\pm$ 11.93 kg; mean $\pm$ standard deviation (SD)). {Data collection for trial 1 was conducted from January to April 2023.} An Intel RealSense 435 depth sensor camera (Intel, Santa Clara, CA, USA) was used to collect videos manually with a resolution of 1280 $\times$ 720 pixels at 60 frames per second. The~camera was installed horizontally above the automatic feeder, 1.51 m above the ground {(Figure~\ref{fig2})}. Video recordings and manual BW measurements were collected on the same day, 2 to 3 times per week, over~4 weeks (28 d) for each animal. Each animal had 8 to 12 pairs of video and BW data. The~total number of video and BW data pairs was 216. Due to technical problems, some videos and BW measurements were lost or unusable, resulting in a reduced dataset. Additionally, one calf died due to a disease unrelated to our research, limiting us to only 4 pairs of video and BW data for that calf. In~total, 196 pairs of BW and corresponding depth videos were included in the further analysis. {Regarding manual BW measurements, obtaining additional BW data would enhance model validation; however, frequent manual weighing can be stressful for calves. Additionally, a~previous study has successfully developed BW prediction models with as few as 12 cows~\cite{bi2023depth}. Therefore, to~balance data sufficiency with animal welfare considerations, we carefully controlled the number of manual weight measurements and expanded the sample size to 20 calves to ensure reliable model training while minimizing handling stress.} 

\begin{table}[H]

\caption{Summary of data collection~trials.\label{tab1}}
\footnotesize
\begin{adjustwidth}{-\extralength}{0cm}
    \begin{tabularx}{\fulllength}{>{\centering}m{1cm}CC>{\centering}m{1.3cm}>{\centering}m{1.7cm}>{\centering}m{1.5cm}C}
        \toprule
        \textbf{Trial} & \textbf{Animals} & \textbf{{{Data} 
 Collection Period}} & \textbf{Video Size} & \textbf{Body Weight Data} & \textbf{Video Data} & \textbf{Purpose} \\
        \midrule
        1 &  Holstein (n = 20); age range: 21--69 days & {January--April 2023} & 196 videos & Manually collected & Manually collected & (1) Correlation between body metrics * and body weight. (2) Body weight prediction using body metrics and actual body~weight. \\
        \midrule
        2 & Holstein (n = 43) and Jersey (n = 5); age range: 14--78 days & {January--June 2024} & 3286 videos & None & Automatically collected & (1) Develop segmentation models using deep learning. (2) Compare segmentation performance between threshold-based and deep learning~methods.\\
        \bottomrule
    \end{tabularx}
\end{adjustwidth}
\noindent{\footnotesize{* Body metrics represent a calf's width, average height, volume, contour area, and~length extracted from images using threshold-based or deep learning segmentation methods.}}
\end{table}

\vspace{-6pt}

\textls[-25]{Trial 2 involved both Holstein (n = 43) and Jersey (n = 5) calves aged 14 to 78 days, with~no BW measurements collected. This imbalance in calf breed distribution reflects the natural population of the farm where data collection occurred. {Data collection for trial 2 occurred from January to June 2024.} A modified automatic recording system, initially developed by \citet{kadlec2022automated} to collect images of cows, was employed. While their system still required human control to turn the camera on and off to collect videos, their automatic feature allowed the camera to detect cows and start recording when cows approached and stop when cows left. Building on this, our system was adapted for dairy calves with further advancements. Our updated system now operates automatically for 20 h a day per day without human intervention, requiring only occasional checks of the camera and computer. It allows the advanced camera settings to automatically adapt to changing light conditions, ensuring optimal performance in the farm’s covered outdoor barn with inconsistent lighting. Our automatic recording system captured 3286 raw videos using the same camera and placed the camera in the same position as trial 1, but~images were captured at 6 frames per second. RealSense SDK tools~\cite{librealsense} were used to convert .bag files into .png images (depth images) and .csv files (depth maps) for both trial~datasets. }

\subsection{Threshold-Based Image Processing for~Segmentation}

\textls[-25]{The threshold segmentation method, applied to each video frame using OpenCV~\cite{bradski2000opencv} for trial 1, converted each frame to HSV (Hue, Saturation, and~Value) color space, extracting the hue channel. Binary thresholding set pixels above 60 to 255 (white) and below 60 to 0 (black). A~threshold of 60 was chosen after testing values from 40 to 70. In~lower thresholds, fence contours surrounding the calf were incorporated into analyses, whereas higher thresholds did not incorporate the full surface area of the calf. The~value of 60 provided the best balance, achieving clear calf segmentation and minimizing fence interference. To~ensure solid contours in each frame, filling was applied to close internal holes within detected objects. Filling is a morphological operation that fills gaps inside an object’s contour, ensuring that the detected shape appears as a solid region rather than a fragmented one. Additionally, opening was used to remove small noise and separate closely positioned objects. Opening is a combination of erosion followed by dilation, which helps eliminate small unwanted artifacts while preserving the overall shape of larger objects. These operations resulted in cleaner and more accurate contours, improving the precision of object detection. For~each video, frames were processed iteratively to detect contours, with~the target contour selected based on specific criteria. First, contours were matched to a template using cv2.matchShapes with a threshold of 0.8 (with smaller values indicating higher similarity, where 0 represents identical shapes). The~template contour was the ideal calf contour that was manually selected from the dataset. This step ensured that detected contours matched the general shape of a calf, filtering out non-calf objects or incomplete shapes. The~0.8 threshold allowed for slight differences in posture or shape while ensuring the contour remains sufficiently similar to the template. Second, contours needed areas between 80,000 and 200,000 pixels (cv2.contourArea). Finally, widths and lengths had to be within 300 to 900 pixels (cv2.boundingRect). This was to avoid including parts of the fence in the contour, given the narrowness of the milk-drinking area. Paired depth .csv files provided depth maps for volume and average height calculations. The~volume was derived from depth pixel sums within the target contour. The~average height was determined by averaging the volume values. The~missing depth data (black pixels) was replaced with the average height. Trial 2 threshold segmentation used the same criteria as trial~1. }

\subsection{Deep Learning Segmentation~Models}

Deep learning model training was conducted on Virginia Tech’s A100 GPU nodes, using an NVIDIA A100-80G GPU, 16 CPU cores, and~247 GB RAM. Pre-trained YOLOv8 models (n, s, m, l, and x) were fine-tuned with a batch size of 16. Training models ran for 100 epochs, with~early stopping after 50 epochs without validation improvement. The~optimizer was set to auto, dynamically adjusting the learning rate. Epoch weights that performed the best were saved to ensure reproducibility. For~trial 2, each video contributed two frames (20-$th$ and 40-$th$ {frame}
), resulting in a total of 6572 images from 3286 videos. {These frames were selected to capture variation in calf positioning while minimizing redundancy in the 6-frame-per-second recordings, ensuring both dataset diversity and storage efficiency.} A semi-supervised approach generated segmented images, starting with 14 manually labeled images in Roboflow (v1.0) \cite{dwyeR2022roboflow} to obtain polygon labels representing the target contour. These labels and images were used to train YOLOv8 models (n, s, m, l, and x) in parallel as baseline models, which then generated pseudo labels for the unlabeled images. Pseudo labels were then manually refined for ground truth creation to improve labeling efficiency and consistency. Instead of fully relying on manual annotations, this iterative adjustment process helped scale up the labeled dataset from 35 calves while reducing human annotation time, expanding it to 55, 122, 509, 1297, and~4794 images. For~evaluation, a~separate set of 957 images from 13 calves was manually labeled. Importantly, neither the animals nor the videos were shared between the training set (35 calves) and the evaluation set (13 calves), ensuring that the model's performance was assessed on completely unseen data. Images were excluded if calves were too close together for clear segmentation or if significant details were missing (e.g., less than half of the calf's contour was visible). These excluded images were removed from the initial set of 6572 raw depth images, leaving 5751 images, which were divided into 4794 for training and 957 for~validation.

\subsection{Segmentation Evaluation Metrics and Statistical~Analysis}
The image segmentation evaluation of threshold and YOLOv8 models (n, s, m, l, and x) used three metrics: intersection over union (IoU), dice coefficient, and~pixel accuracy, using 957 images from trial 2. The~polygon ground truth labels (which outline objects with lines) of 957 images were converted into the corresponding ground truth masks (solid filled regions) using the cv2.fillPoly function for segmentation evaluation. These metrics evaluated the deep learning models or threshold-based segmentation method by comparing the segmented masks with the ground truth masks. Intersection over union assessed overlap by comparing the intersection and union of segmented and ground truth masks. The~dice coefficient measured segmentation similarity, using a ratio of twice the pixel intersection to the total pixel area in both masks. Pixel accuracy calculated the percentage of correctly classified pixels by dividing the number of correctly classified pixels by the total pixels. Each metric ranges from 0 to 1, with~higher values indicating better segmentation accuracy. An~Analysis of Variance (ANOVA) tested performance differences across segmentation methods. A~post hoc analysis was conducted using Tukey's Honest Significant Difference (HSD) to account for multiple comparisons. A~\textit{p} value < 0.05 was considered biologically significant. {Eta squared ($\eta^2$) was calculated as an effect size to quantify the proportion of variance in segmentation performance explained by the choice of segmentation method.}

\subsection{Correlation Analysis of Body Metrics and Actual Body~Weight}

Correlations among body metrics (contour area, volume, average height, width, and~length) were assessed using threshold and YOLOv8 models (n, s, m, l, and~x) with actual BW measurements from trial 1. Since each calf had multiple images per date, the~median body metrics for each calf by date were used for correlation analyses. These metrics were used in further BW prediction analysis. Pearson correlation graphs visualized these relationships.  To~further understand the relationships between body metrics and BW across different age groups, we divided the dataset into quartiles based on the age of the calves, focusing on body metrics obtained from the YOLOv8m segmentation method. The~model YOLOv8m was selected due to its slightly high segmentation performance and its balance between accuracy and computational efficiency. Correlation matrices among age groups were compared using a Mantel test to assess consistency in YOLOv8m prediction metrics and actual BW. A~Mantel test result with \textit{p} < 0.05 indicated similar correlation structures between age~groups.

\subsection{Body Weight~Prediction}
\subsubsection{Cross-Validation for Single-Time-Point Body Weight~Prediction}
\textls[-25]{For predicting BW at a single time point, body metrics were first extracted using segmentation approaches: deep learning-based segmentation with YOLOv8 models (n, s, m, l, x) and the threshold-based segmentation method. These extracted body metrics, along with actual BW from trial 1, were then used to train and evaluate LR and XGBoost models for BW prediction. A~5-fold cross-validation approach was implemented to prevent data leakage, ensuring no overlap of calves or images between the training set (4 folds; 16 calves) and the testing set (1 fold; 4 calves). This approach helps the models learn from different subsets of data, improving models' ability to make accurate predictions on new, unseen calves rather than just memorizing patterns from the training set. To~further assess robustness, single-time-point cross-validation was repeated 100 times with randomized calf reassignments, providing diverse data splits. After~these 100 repetitions, performance metrics such as R$^{2}$, mean squared error (MSE), root mean squared error (RMSE), mean absolute error (MAE), and~mean absolute percentage error (MAPE) were calculated and averaged with SD, ensuring a robust assessment of model accuracy while accounting for variability in the dataset. An~ANOVA tested performance differences between LR and XGBoost for each metric (R$^{2}$, MSE, RMSE, MAE, and MAPE). A~\textit{p} value < 0.05 was considered biologically~significant.}

For LR, the~standard LR model follows the general form as described in~\cite{montgomery2021introduction}. In~this study, BW was modeled as a function of fixed effects, including age, length, width, average height, volume, and~contour area, to~capture their contributions to BW prediction. For~XGBoost, a~randomized search was used to optimize model hyperparameters, including the learning rate (0.01 to 0.9), the~number of estimators (50 to 10,000), and~the regularization terms alpha and lambda (both ranging from 0 to 1). Within~these specified ranges, the~randomized search algorithm iteratively samples combinations of hyperparameters (with 1000 iterations) to identify the optimal set that minimizes error and maximizes model performance. The~models were implemented in Python 3.9 using the XGBoost package (version 1.5.2) and Scikit-learn (version 1.2.2) for~LR.

\subsubsection{Longitudinal Analysis for Multiple-Time-Point Body Weight~Prediction}

\textls[-25]{Using the same data as in the single-time-point cross-validation analysis to predict BW, the~difference is that the forecasting was conducted with different train--test splits (\(90:10\), \(80:20\), \(70:30\), \(60:40\), and~\(50:50\)) using LR, XGBoost, and~LMM prediction models. Each calf's BW was recorded at multiple time points manually, and~the dataset was split based on time rather than randomly. The~first portion (\(X\%\)) of each calf’s time-series data was used for training, while the remaining (\(100-X\%\)) was reserved for testing. X are 90, 80, 70, 60, and~50, respectively. For~data preprocessing, one calf with only four recorded data points was removed to ensure sufficient longitudinal representation for model training. To~simulate real-world scenarios with missing data, a~randomly selected calf at one random time point was excluded before the train--test split. This process was repeated 100 times to account for variability in missing data scenarios and to ensure robust evaluation. Additionally, multiple iterations were conducted to generate the SD of performance metrics, which were then used for statistical comparison, as~longitudinal analysis is inherently performed only once. An~ANOVA was conducted to compare the performance of the three predictive models (LR, XGBoost, and~LMM) across different train--test splits. Model performance was evaluated using R$^{2}$, MSE, RMSE, MAE, and~MAPE. A~post hoc analysis was conducted using Tukey's HSD test to account for multiple comparisons. A~\textit{p} value < 0.05 was considered biologically significant. {Eta squared ($\eta^2$) was calculated to quantify the effect size, providing insight into the proportion of variance explained by the choice of BW predictive model.}}

\textls[-25]{For the modeling approach, LR, XGBoost, and~LMM were used to predict BW over time. Linear regressions followed the same modeling strategy as in the single-time-point cross-validation analysis, maintaining the same predictor variables and fixed effects, while XGBoost retained the same hyperparameter tuning approach. However, unlike LR and XGBoost, LMM was introduced to account for repeated measurements within calves. The~standard LMM equation is described in \citet{pinheiro2006mixed}. In~this study, LMM incorporated fixed effects (age, length, width, average height, volume, and~contour area) and a random effect for calf ID to model individual growth trajectories while capturing within-individual variation over time. The LMM was implemented using the lme4 package (version 1.1.36) in R 4.3.0~\cite{bates2015fitting}.}

\section{Results}
\subsection{Segmentation Performance~Evaluation}
\begin{figure}[H]
\begin{adjustwidth}{-\extralength}{0cm}
\centering
\includegraphics[width=15.5cm]{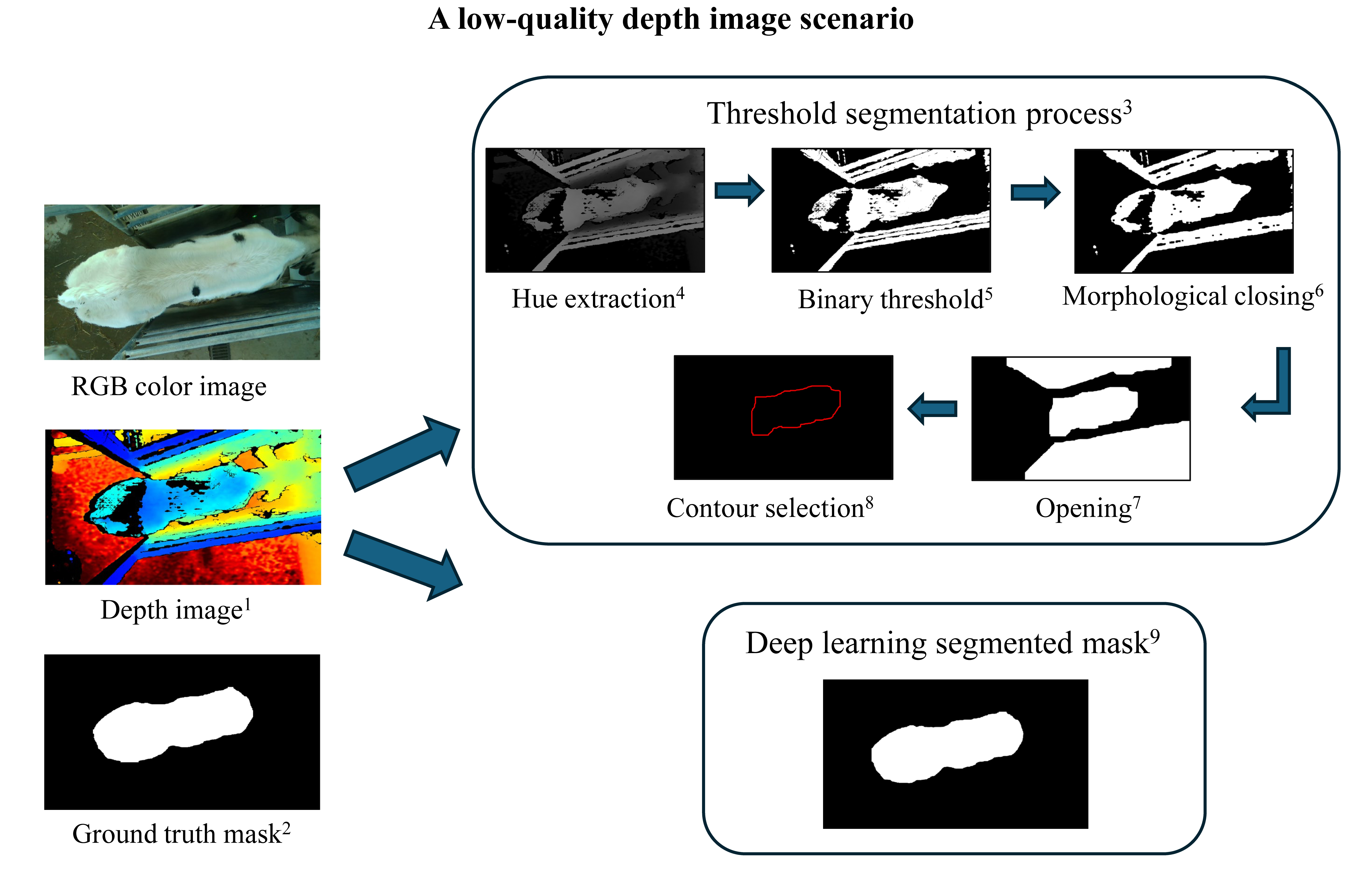}
\end{adjustwidth}
\caption{A low-quality depth image scenario and segmentation process. Threshold-based segmentation failed to generate the ideal calf contour due to the low-quality depth image and overexposure on the white pattern of the calf, while the deep learning model successfully segmented the calf contour with higher {accuracy.} 
\textsuperscript{1} Depth image: an image where pixel values represent the distance between the camera and objects in the~scene. 
\textsuperscript{2} Ground truth mask: a manually annotated segmentation mask used as a reference to evaluate the accuracy of segmentation~methods. 
\textsuperscript{3} Threshold segmentation process: a rule-based method that involves extracting specific pixel intensity values to differentiate objects from the~background. 
\textsuperscript{4} Hue extraction: converts the image into HSV (hue, saturation, value) color space and extracts the hue channel, which represents color information independent of~brightness.
\textsuperscript{5} Binary threshold: converts the grayscale image into a binary image by setting pixel values above a chosen threshold (60) to 255 (white, object) and those below it to 0 (black, background).
\textsuperscript{6} Morphological closing: a morphological operation that applies dilation followed by erosion to close small holes or gaps within detected objects, ensuring solid~contours.
\textsuperscript{7} Opening: a morphological operation that applies erosion followed by dilation to remove small noise and separate closely positioned~objects.
\textsuperscript{8} Contour selection: identifies object boundaries in the segmented image. The~target contour is chosen based on its similarity to a predefined calf shape and constraints on area, width, and~length.
\textsuperscript{9} Deep learning segmented mask: a segmentation mask generated by a trained YOLOv8 model to recognize the calf from the depth~image.
\label{fig3}}
\end{figure}

\begin{figure}[H]
\includegraphics[width=13.5cm]{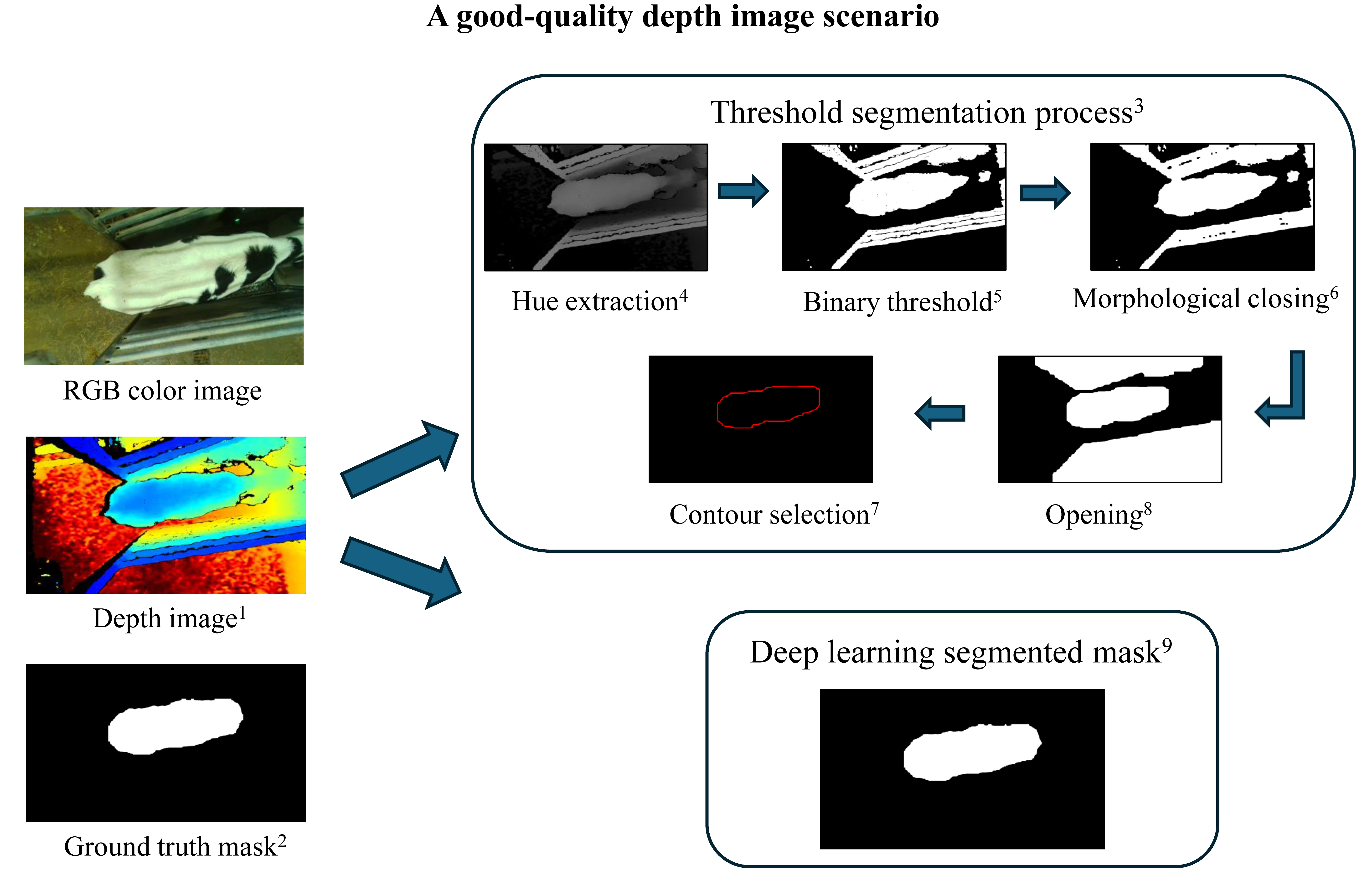}
\caption{ A good-quality depth image scenario and segmentation process. Threshold-based segmentation and the deep learning model successfully extracted the calf contour from the good-quality depth image, achieving accurate segmentation results. The~definitions from \textsuperscript{1--9} are the same as described in Figure~\ref{fig3}.\label{fig4}}
\end{figure}

The threshold method segmented 573 images (60\%) out of the original 957 from trial 2, with~the remaining 40\% failing due to over- or underexposure and/or area of image collection. Calves leaning against the fence while drinking during image collection reduced contour detection. An~example of a failed segmentation is shown in Figure~\ref{fig3}, while a good-quality image scenario is presented in Figure~\ref{fig4}.  In~contrast, all YOLOv8 models (n, s, m, l, and x) segmented 100\% of the images, demonstrating robustness to lighting and complex environments (Table~\ref{tab2}). {Furthermore, ANOVA results indicated that segmentation method choice had a statistically significant effect on segmentation performance (adjusted \textit{p} = $4.47 \times 10^{-149}$ for IoU, adjusted \textit{p} = $3.15 \times 10^{-102}$ for dice coefficient, and~adjusted \textit{p} = $3.53 \times 10^{-71}$ for pixel accuracy; Table~\ref{tab2}). These extremely small adjusted \textit{p}-values highlighted the strong impact of segmentation method choice on performance, emphasizing the importance of selecting an optimal method. Additionally, effect size analysis using eta squared ($\eta^2$) showed that segmentation method choice explained 12.3\% of the variation in IoU, 8.7\% in dice coefficient, and~6.2\% in pixel accuracy. The~higher effect size for IoU suggested a stronger influence of segmentation method, likely due to the threshold-based method’s significantly lower IoU (0.888) compared to YOLOv8 models (IoU $\geq 0.96$). In~contrast, dice coefficient and pixel accuracy had smaller differences, leading to lower effect sizes. These results indicated that IoU is more sensitive to the choice of the segmentation method in our study, emphasizing its importance in evaluating segmentation performance.} Moreover, pairwise comparisons revealed significant segmentation performance differences between YOLOv8 and threshold methods (\textit{p} < 0.001), with~YOLOv8m, YOLOv8l, and~YOLOv8x achieving increased accuracy compared to YOLOv8s and YOLOv8n models (\textit{p} < 0.05) (Table~\ref{tab2}). The~model YOLOv8m was selected for correlation and BW prediction result visualization due to its slightly high segmentation performance and its balance between accuracy and computational~efficiency.

\begin{table}[H]
\scriptsize
\caption{Comparison of segmentation performance (IoU $^{1}$, dice coefficient $^{2}$, pixel accuracy $^{3}$) between the threshold-based segmentation method and YOLOv8 segmentation models. Values are presented as mean $\pm$ standard~deviation {(SD)}.\label{tab2}}
\begin{adjustwidth}{-\extralength}{0cm}
    \newcolumntype{C}{>{\centering\arraybackslash}X}
    \begin{tabularx}{\fulllength}{>{\centering}m{2cm}CCCCCCCCC}
            \toprule
        \textbf{Metrics} & \textbf{YOLOv8n} & \textbf{YOLOv8s} & \textbf{YOLOv8m} & \textbf{YOLOv8l} & \textbf{YOLOv8x} & \textbf{Threshold} & {\textbf{\textit{p}-Value} \textsuperscript{\textbf{5}}} & {\textbf{Adjusted \textit{p}-Value} \textsuperscript{\textbf{6}}} & {\textbf{Eta Squared (\boldmath{$\eta^2$}) \textsuperscript{\textbf{7}}}} \\

        \midrule
        Segmented Image Numbers (Percentage) $^{4}$ & 957 (100\%) & 957 (100\%) & 957 (100\%) & 957 (100\%) & 957 (100\%) & 573 (60\%) & {-} & {-} & {-} \\
        IoU: \% {Mean ± SD} & 0.966 $\pm$ {0.07} 
 $^{b,c}$ & 0.96 $\pm$ 0.07 $^{c}$ & 0.976 $\pm$ 0.06 $^{a}$ & 0.973 $\pm$ 0.06 $^{a,b}$ & 0.975 $\pm$ 0.06 $^{a,b}$ & 0.888 $\pm$ 0.09 $^{d}$ & {$7.46 \times 10^{-150}$} & {$4.47 \times 10^{-149}$} & {0.123} \\
        Dice Coefficient: \% {Mean ± SD} & 0.981 $\pm$ 0.04 $^{b,c}$ & 0.978 $\pm$ 0.05 $^{c}$ & 0.987 $\pm$ 0.04 $^{a}$ & 0.986 $\pm$ 0.04 $^{a,b,c}$ & 0.986 $\pm$ 0.04 $^{a,b}$ & 0.938 $\pm$ 0.06 $^{d}$ & {$5.25 \times 10^{-103}$} & {$3.15 \times 10^{-102}$} & {0.087} \\
        Pixel Accuracy: \% {Mean ± SD}
        & 0.993 $\pm$ 0.02 $^{b}$ & 0.992 $\pm$ 0.02 $^{c}$ & 0.995 $\pm$ 0.02 $^{a}$ & 0.995 $\pm$ 0.02 $^{a,b}$ & 0.995 $\pm$ 0.02 $^{a,b}$ & 0.98 $\pm$ 0.02 $^{d}$ & {$5.88 \times 10^{-72}$} & {$3.53 \times 10^{-71}$} & {0.062} \\
        
        \bottomrule
    \end{tabularx}
\end{adjustwidth}

\noindent{\footnotesize{
{\textsuperscript{1}} IoU: intersection over union, a~metric used to evaluate the accuracy of an object segmentation model by measuring the ratio of the overlap between the predicted and ground truth regions to the combined area of both
 regions, with~values ranging from 0 (no~ overlap) to 1 (perfect overlap). \textsuperscript{2} Dice coefficient: measures the similarity between two~ segmentation masks. It is~ calculated as twice~ the area of overlap between the predicted and ground truth regions divided by the total number of pixels in both regions, with values ranging from 0 (no overlap) to 1 (perfect overlap). \textsuperscript{3} Pixel~ accuracy: ~~measures~~ the percentage~~ of~~ correctly   
classified pixels in an image, with~values ranging from 0 (no correct pixels) to 1 (all pixels correctly classified). \textsuperscript{4} Here, 957 images from 13 calves were used for segmentation evaluation from trial 2; the numbers represent the total images where the calf contour was successfully segmented, while the values in parentheses indicate the percentage of successfully segmented~images. 
{\textsuperscript{5} \textit{p}-values indicate the statistical significance of differences among segmentation methods.} 
{\textsuperscript{6} Adjusted \textit{p}-values were obtained using the Bonferroni correction to control for multiple comparisons, ensuring that statistical significance is maintained while reducing the risk of Type I errors.} 
{\textsuperscript{7} Eta squared ($\eta^2$) is an effect size measure that quantifies the proportion of variation in segmentation performance attributed to the choice of segmentation method. Higher $\eta^2$ values indicate a greater influence of the method on performance.} 
{\textit{\textsuperscript{a--d}}} 
The segmentation performance in the same row indicates statistically significant differences (\textit{p} < 0.05) by applying an Analysis of Variance (ANOVA) followed by Tukey's Honest Significant Difference (HSD) post hoc test. Methods sharing the same letter (a, b, c, or~d) are not significantly different, while those with different letters show significant differences.
}}
\end{table}

\vspace{-6pt}

\vspace{-6pt}

\unskip

\subsection{Correlation Analysis of Body Metrics and Actual Body~Weight}
\begin{figure}[H]
\includegraphics[width=13.5 cm]{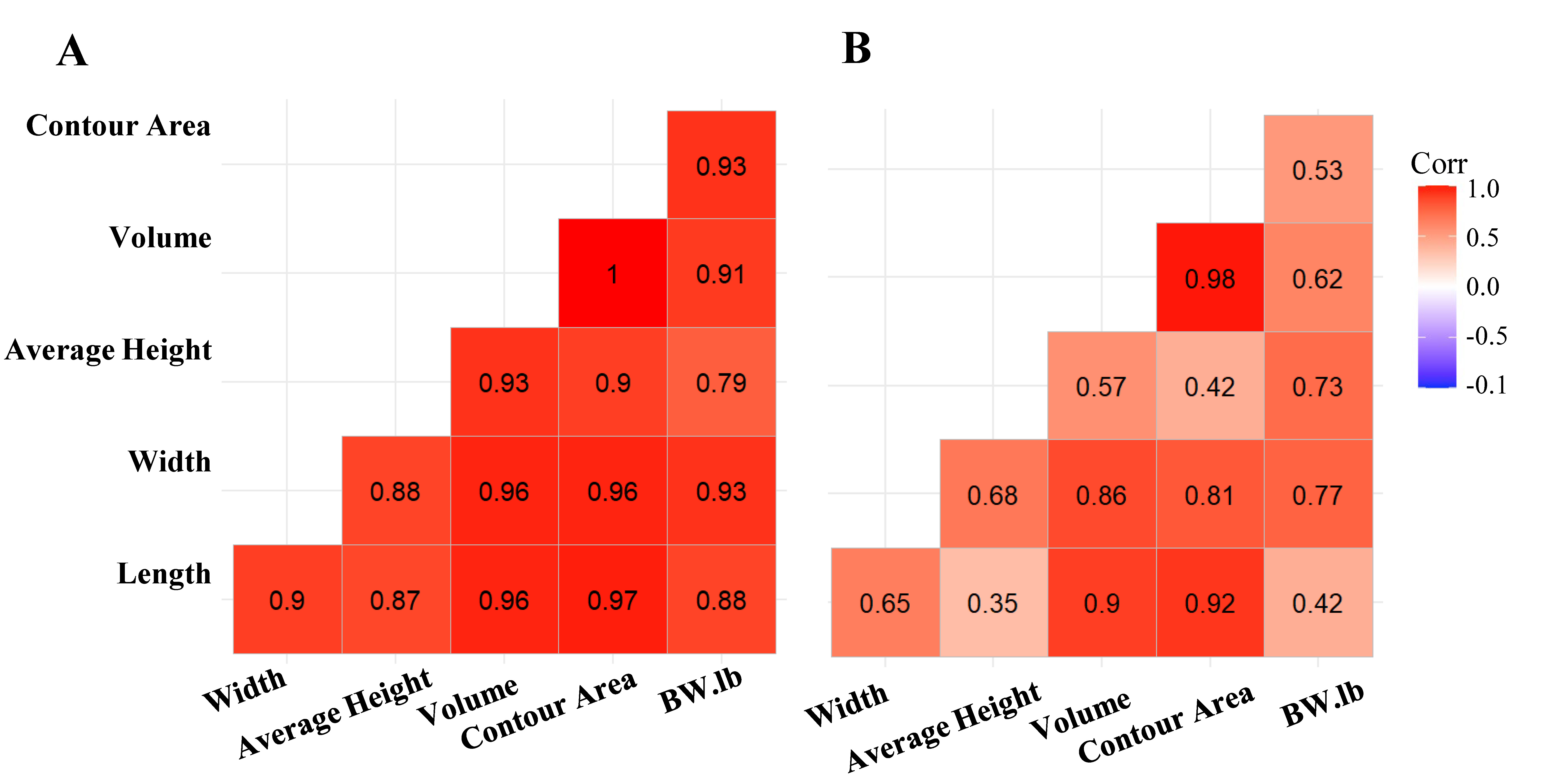}
\caption{{Pearson} 
 correlation matrices for body metrics (width, average height, volume, contour area, and~length) and actual body weight using (\textbf{A}) YOLOv8m and (\textbf{B}) threshold segmentation. Darker red shades indicate stronger correlations, with~YOLOv8m demonstrating increased correlations with body~weight.\label{fig5}}
\end{figure}

Using the threshold segmentation method, 191 of 196 depth videos were segmented to obtain the calf contour for measuring the body metrics from trial 1. In~comparison, all 196 depth videos were segmented using YOLOv8 (n, s, m, l, and x). Since trial 1 data were collected manually during non-feeding times, calves were often in unnatural positions, such as with their heads raised or in varied postures. Additionally, some images were overexposed due to the reflective nature of white coat color patterns. These factors made it difficult for the threshold segmentation method to accurately detect body contours, leading to the failure of threshold segmentation in a few images. Figure~\ref{fig5}B illustrates moderate correlations between BW and prediction metrics such as average height and volume. There was a positive correlation between BW and the prediction body metric width (0.77) and height (0.73). These findings suggest that the threshold method moderately captured BW and body metric relationships. In~contrast, Figure~\ref{fig5}A presents that the deep learning approach (YOLOv8m) yielded stronger correlations with BW, with~values of 0.91 for volume and 0.93 for contour area. This indicates that the deep learning method more effectively extracted features that correlated with calf BW, leading to more accurate~predictions.

\unskip  

\subsection{Age-Based Correlation Analysis of Body Metrics and Actual Body~Weight}
\begin{figure}[H]
\includegraphics[width=10.5 cm]{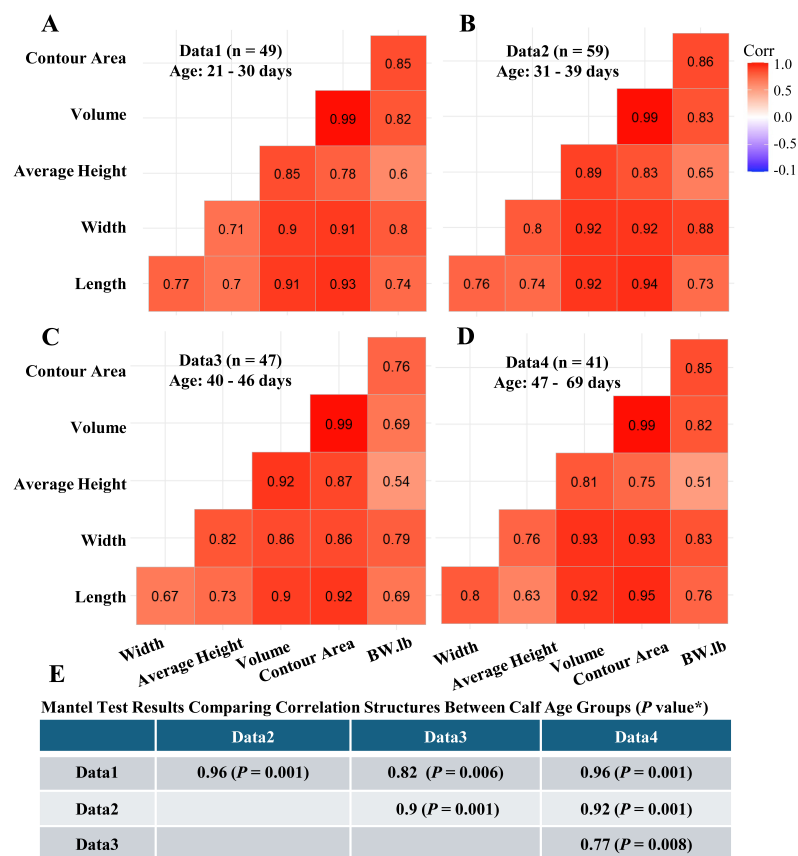}
\caption{{Pearson} 
 correlation matrices for body metrics (width, average height, volume, contour area, and~length) and actual body weight across different calf age groups. The~body metrics were extracted using the YOLOv8 segmentation model. Panels (A--D) show correlation matrices for calves aged (\textbf{A}) 21--30 days, (\textbf{B}) 31--39 days, (\textbf{C}) 40--46 days, and~(\textbf{D}) 47--69 days. The~\textbf{n} in each figure indicates the number of pairs of body metrics and body weight data points included in each dataset. The~correlations are visualized to assess how the relationship between these metrics and body weight changes as the calves mature. Panel (\textbf{E}) presents the results of Mantel tests comparing the similarity of these correlation matrices across the four age~groups.\label{fig6}}
\end{figure}

Pearson correlation matrices for various body metrics and BW across calf age were derived from the YOLOv8m deep learning segmentation method (Figure~\ref{fig6}A--D). The~dataset (n = 196 pairs of body metric and BW; 20 calves; trail 1) was divided into age quartiles, generating correlation matrices for each group: 21--30 days (Figure~\ref{fig6}A), 31--39 days (Figure~\ref{fig6}B), 40--46 days (Figure~\ref{fig6}C), and~47--69 days (Figure~\ref{fig6}D). Strong correlations between body metrics and BW across all age groups indicated robustness in these relationships throughout calf growth. Mantel tests (Figure~\ref{fig6}E) showed correlation structure similarities among age groups (\textit{p} values = 0.001 to 0.008), suggesting consistent relationships between body metrics and BW as calves age. {Thus, body metrics measured at early time points may be used to predict body weight at later time points.} Furthermore, these correlation findings informed the subsequent BW prediction analysis, which involved forecasting cross-validation (using early time point measurements to predict BW at later ages) to evaluate prediction accuracy across different segmentation~approaches.

\unskip

\subsection{Body Weight~Prediction}
\subsubsection{Cross-Validation for Single-Time-Point Body Weight~Prediction}  
A five-fold cross-validation approach was used to train and evaluate LR and XGBoost models, using body metrics and actual BW as input features. Table~\ref{tab3} presents the performance of LR and XGBoost for BW prediction across two segmentation methods: YOLOv8m and threshold-based segmentation. Across all metrics (R$^{2}$, MSE, RMSE, MAE, and MAPE), XGBoost significantly outperformed LR {(\textit{p} < 0.001; Table~\ref{tab3}), as~indicated by the superscripts \( a, b \)}, with~YOLOv8m achieving the highest R$^{2}$ (0.91) and the lowest error values {(\textit{p} < 0.001; Table~\ref{tab3})}. In~contrast, the~threshold-based approach resulted in lower prediction accuracy, with~R$^{2}$ values of 0.73 for LR and 0.8 for XGBoost, along with higher error metrics {(\textit{p} < 0.001; Table~\ref{tab3})}. This suggests that XGBoost was better able to leverage the higher-quality body metrics extracted from YOLOv8m, capturing more informative patterns for BW prediction, whereas its performance was somewhat limited by the less precise features obtained through threshold-based segmentation. Results for other YOLO segmentation models (n, s, l, and x) and BW predictions using LR and XGBoost are provided in Table~S1.

\begin{table}[H]
\caption{Comparison of body weight prediction performance using linear regression (LR) and extreme gradient boosting (XGBoost) models with single-time-point cross-validation across YOLOv8m and threshold-based segmentation methods. Values are presented as mean $\pm$ standard deviation (SD) \textsuperscript{1}.\label{tab3}}
	\begin{adjustwidth}{-\extralength}{0cm}
		\begin{tabularx}{\fulllength}{CCCCCCC}
			\toprule
			\multirow{2}{*}{\textbf{Metric \textsuperscript{2}}} & \multicolumn{3}{c}{\textbf{YOLOv8m}} & \multicolumn{3}{c}{\textbf{Threshold}} \\
			\cmidrule{2-7} 
			            & \textbf{LR} & \textbf{XGBoost} & \textbf{{\textit{p}-Value \boldmath{$^{3}$}}} & \textbf{LR} & \textbf{XGBoost} & \textbf{{\textit{p}-Value \boldmath{$^{4}$}}} \\
		      		      \cmidrule{1-7} 
			R$^{2}$        & 0.85 $\pm$ 0.09 $^{b}$  & {0.91 $\pm$ 10$^{-3.5}$ $^a$}  & {8.80 $\times$ 10$^{-12}$} & 0.73 $\pm$ 0.09 $^{b}$  & {0.8 $\pm$ 10$^{-3.5}$ $^a$} & {3.00 $\times$ 10$^{-14}$}  \\
			MSE (lb$^{2}$) & 82.24 $\pm$ 36.15 $^{a}$  & 61.74 $\pm$ 0.23 $^{b}$   & {6.89 $\times$ 10$^{-8}$}  & 168.38 $\pm$ 47.7 $^{a}$  & 133.31 $\pm$ 0.31 $^{b}$ & {2.00 $\times$ 10$^{-12}$} \\
			RMSE (lb) & 8.81 $\pm$ 2.16 $^{a}$  & 7.86 $\pm$ 0.01 $^{b}$   & {4.69 $\times$ 10$^{-6}$}  & 12.84 $\pm$ 1.87 $^{a}$  & 11.55 $\pm$ 0.01 $^{b}$  & {2.75 $\times$ 10$^{-10}$}  \\
			MAE (lb)  & 7.21 $\pm$ 1.88 $^{a}$  & 6.22 $\pm$ 0.01 $^{b}$   & {1.39 $\times$ 10$^{-8}$}  & 10.56 $\pm$ 1.71 $^{a}$  & 9.47 $\pm$ 0.01 $^{b}$  & {7.48 $\times$ 10$^{-8}$}  \\
			MAPE (\%)  & 5.03 $\pm$ 1.07 $^{a}$  & 4.37 $\pm$ 0.01 $^{b}$   & {8.80 $\times$ 10$^{-8}$}  & 7.4 $\pm$ 1.39 $^{a}$  & 6.73 $\pm$ 0.01 $^{b}$  & {5.69 $\times$ 10$^{-8}$}  \\

		\bottomrule
		\end{tabularx}
	\end{adjustwidth}
\noindent{\footnotesize{%
\textsuperscript{1} Mean and SD were computed across 100 cross-validation repetitions, where each repetition involved a new randomized split of training (16 calves) and testing (4 calves), ensuring robustness in model performance~assessment. 
\textsuperscript{2} R$^{2}$ = coefficient of determination, MSE = mean squared error, RMSE = root mean square error, MAE = mean absolute error, and MAPE = mean absolute percentage~error.
{\textsuperscript{3} \textit{p}-value was derived from an ANOVA test. For~the YOLOv8m segmentation method, the~\textit{p}-value evaluates the statistical significance of the performance differences between the LR and XGBoost models, with~\textit{p} < 0.05 indicating a statistically significant~difference. 
\textsuperscript{4} \textit{p}-value was derived from an ANOVA test. 
For the threshold-based segmentation method, the~\textit{p}-value assesses the statistical difference in performance between LR and XGBoost, with~\textit{p} < 0.05 indicating a statistically significant difference. }
$^{a,b}$ superscripts compared the performance of LR and XGBoost models for each metric (R$^{2}$, MAE, MAPE, MSE, and RMSE). Different superscripts indicate significant differences (\textit{p} < 0.05) based on the Analysis of Variance (ANOVA) test. For~each segmentation method (YOLOv8m and threshold), superscripts compared the performance of LR and XGBoost within the same method. That is, within~YOLOv8m, LR and XGBoost were compared, and~separately, within~threshold, LR and XGBoost were compared.}}
\end{table}
\unskip

\subsubsection{Longitudinal Analysis for Multiple-Time-Point Body Weight~Prediction}
Using a time-series approach, BWs were predicted by using LR, XGBoost, and~LMM across different train--test splits (\(90:10\), \(80:20\), \(70:30\), \(60:40\), and~\(50:50\)). 
 The~results demonstrated that LMM consistently outperformed LR and XGBoost across all models and train--test splits, achieving the highest R$^{2}$ values and the lowest errors (MSE, RMSE, MAE, and~MAPE), indicating strong predictive accuracy. {The extremely small adjusted \textit{p}-values indicated that the differences between the BW prediction models were statistically significant, confirming that model choice plays a crucial role in predictive performance.  Furthermore, the~eta squared ($\eta^2$) values were all greater than 0.95, demonstrating a large effect size, meaning that the choice of predictive model explains a substantial proportion of the variance in BW prediction accuracy (Tables~\ref{tab4} and \ref{tab5}).}

When using YOLOv8m segmentation, all three models achieved better predictive performance compared to the threshold-based approach. The~LMM approach exhibited near-perfect predictions, with an R$^{2}$ of 0.99 in the 90:10 split and 0.92 in the 50:50 split, with~significantly lower error values, demonstrating its robustness even as training data decreased. The~LR performed better than XGBoost in YOLOv8m segmentation, with~consistently lower error rates and a higher R$^{2}$, indicating that it was more effective at modeling the relationships between body metrics and BW. The~XGBoost had slightly lower R$^{2}$ values and higher errors compared to LR, suggesting increased variability in its predictions. However, LR and XGBoost showed relatively stable performance across different train--test splits, with~LR achieving an R$^{2}$ of 0.89 in the 50:50 split, while XGBoost dropped to 0.82 ({Table ~\ref{tab4}}). 

\begin{table}[H]
\caption{Comparison of body weight prediction performance using linear regression (LR), extreme gradient boosting (XGBoost), and~linear mixed models (LMM) with cross-validation {using YOLOv8m segmentation}. Values are presented as mean $\pm$ standard deviation (SD){\textsuperscript{1}}. \label{tab4}}
	\begin{adjustwidth}{-\extralength}{0cm}
		\begin{tabularx}{\linewidth}{>{\centering\arraybackslash}p{0.075\linewidth} >{\centering\arraybackslash}p{0.06\linewidth} >{\centering\arraybackslash}p{0.12\linewidth} >{\centering\arraybackslash}p{0.12\linewidth} >{\centering\arraybackslash}p{0.12\linewidth} >{\centering\arraybackslash}p{0.12\linewidth} >{\centering\arraybackslash}p{0.12\linewidth} >{\centering\arraybackslash}p{0.07\linewidth}}
		  \toprule
			\textbf{Metric{\textsuperscript{2}}} & \textbf{Train:Test{\textsuperscript{3}}} & \textbf{LR} & \textbf{XGBoost} & \textbf{LMM} & \textbf{{\textit{p}-Value}{\textsuperscript{4}}} & \textbf{{Adjusted \textit{p}-Value}{\textsuperscript{5}}} & \textbf{{{Eta Squared (\boldmath{$\eta^2$}){\textsuperscript{6}}}}} \\
			\midrule
            \multirow{5}{*}{{R$^2$}}  
                & 90:10 & {0.90 $\pm$ 0.009 $^b$}  & {0.89 $\pm$ 0.003 $^c$}  & {0.99 $\pm$ 0.002 $^a$} & {1.42 $\times$ 10$^{-283}$} & {4.25 $\times$ 10$^{-283}$} & {0.988} \\
                & 80:20 & {0.90 $\pm$ 0.005 $^b$}  & {0.88 $\pm$ 0.002 $^c$}  & {0.96 $\pm$ 0.001 $^a$} & {Inf $\times$ 10$^{-324}$}  & {1.5 $\times$ 10$^{-323}$}  & {0.993} \\
                & 70:30 & {0.91 $\pm$ 0.004 $^b$}  & {0.89 $\pm$ 0.002 $^c$}  & {0.95 $\pm$ 0.003 $^a$} & {2.3 $\times$ 10$^{-287}$} & {6.9 $\times$ 10$^{-287}$} & {0.988} \\
               & 60:40 & {0.91 $\pm$ 0.002 $^b$}  & {0.88 $\pm$ 0.01 $^c$}  & {0.95 $\pm$ 0.006 $^a$} & {4.3 $\times$ 10$^{-256}$} & {1.29 $\times$ 10$^{-255}$} & {0.981} \\
                & 50:50 & {0.89 $\pm$ 0.005 $^b$}  & {0.82 $\pm$ 0.01 $^c$}  & {0.92 $\pm$ 0.01 $^a$} & {1.42 $\times$ 10$^{-250}$} & {4.25 $\times$ 10$^{-250}$} & {0.979} \\
            \midrule
            \multirow{5}{*}{MSE (lb$^2$)}  
                & 90:10 & 46.79 $\pm$ 1.73 $^b$  & 75.07 $\pm$ 2.56 $^a$  & 19.54 $\pm$ 0.72 $^c$  & {<1 $\times$ 10$^{-300}$} & {<1 $\times$ 10$^{-300}$} & {0.994} \\
                & 80:20 & 71.65 $\pm$ 1.59 $^b$  & 77.51 $\pm$ 1.11 $^a$  & 22.31 $\pm$ 0.58 $^c$  & {<1 $\times$ 10$^{-300}$} & {<1 $\times$ 10$^{-300}$} & {0.998} \\
                & 70:30 & 82.22 $\pm$ 1.15 $^a$  & 71.12 $\pm$ 1.02 $^b$  & 26.84 $\pm$ 0.98 $^c$  & {<1 $\times$ 10$^{-300}$} & {<1 $\times$ 10$^{-300}$} & {0.998} \\
                & 60:40 & 76.45 $\pm$ 0.94 $^a$  & 76.29 $\pm$ 4 $^a$  & 34.01 $\pm$ 1.34 $^b$  & {1.88 $\times$ 10$^{-274}$} & {5.63 $\times$ 10$^{-274}$} & {0.986} \\
                & 50:50 & 89.69 $\pm$ 2 $^b$  & 112.45 $\pm$ 5.17 $^a$  & 55.77 $\pm$ 3.29 $^c$  & {4.65 $\times$ 10$^{-243}$} & {1.39 $\times$ 10$^{-242}$} & {0.977} \\
            \midrule
            \multirow{5}{*}{RMSE (lb)}  
                & 90:10 & 6.84 $\pm$ 0.12 $^b$  & 8.66 $\pm$ 0.15 $^a$  & 4.42 $\pm$ 0.08 $^c$  & {<1 $\times$ 10$^{-300}$} & {<1 $\times$ 10$^{-300}$} & {0.996} \\
                & 80:20 & 8.46 $\pm$ 0.1 $^b$  & 8.8 $\pm$ 0.06 $^a$  & 4.72 $\pm$ 0.06 $^c$  & {<1 $\times$ 10$^{-300}$} & {<1 $\times$ 10$^{-300}$} & {0.999} \\
                & 70:30 & 9.07 $\pm$ 0.06 $^a$  & 8.43 $\pm$ 0.06 $^b$  & 5.18 $\pm$ 0.09 $^c$  & {<1 $\times$ 10$^{-300}$} & {<1 $\times$ 10$^{-300}$} & {0.998} \\
                & 60:40 & 8.74 $\pm$ 0.05 $^a$  & 8.73 $\pm$ 0.22 $^a$  & 5.83 $\pm$ 0.11 $^b$  & {1.02 $\times$ 10$^{-296}$} & {3.05 $\times$ 10$^{-296}$} & {0.990} \\
                & 50:50 & 9.47 $\pm$ 0.11 $^b$  & 10.6 $\pm$ 0.24 $^a$  & 7.46 $\pm$ 0.23 $^c$  & {2.02 $\times$ 10$^{-248}$} & {6.07 $\times$ 10$^{-248}$} & {0.979} \\
                \multirow{5}{*}{MAE (lb)}  
                & 90:10 & 5.01 $\pm$ 0.14 $^b$  & 7.11 $\pm$ 0.15 $^a$  & 3.90 $\pm$ 0.10 $^c$  & {5.48 $\times$ 10$^{-307}$} & {1.64 $\times$ 10$^{-306}$} & {0.991} \\
                & 80:20 & 6.85 $\pm$ 0.08 $^b$  & 7.14 $\pm$ 0.07 $^a$  & 3.76 $\pm$ 0.06 $^c$  & {<1 $\times$ 10$^{-300}$}  & {<1 $\times$ 10$^{-300}$}  & {0.998} \\
                & 70:30 & 7.50 $\pm$ 0.07 $^a$  & 6.73 $\pm$ 0.07 $^b$  & 4.11 $\pm$ 0.09 $^c$  & {<1 $\times$ 10$^{-300}$}  & {<1 $\times$ 10$^{-300}$}  & {0.998} \\
                & 60:40 & 7.11 $\pm$ 0.06 $^a$  & 7.06 $\pm$ 0.24 $^b$  & 4.48 $\pm$ 0.10 $^c$  & {9.82 $\times$ 10$^{-272}$}  & {2.94 $\times$ 10$^{-271}$}  & {0.985} \\
                & 50:50 & 7.82 $\pm$ 0.10 $^b$  & 8.51 $\pm$ 0.27 $^a$  & 5.73 $\pm$ 0.16 $^c$  & {5.36 $\times$ 10$^{-241}$}  & {1.61 $\times$ 10$^{-240}$}  & {0.976} \\
            \midrule
            \multirow{5}{*}{MAPE (\%)}  
                & 90:10 & 3.17 $\pm$ 0.10 $^b$  & 4.60 $\pm$ 0.10 $^a$  & 2.39 $\pm$ 0.06 $^c$  & {7.03 $\times$ 10$^{-309}$}  & {2.11 $\times$ 10$^{-308}$}  & {0.992} \\
                & 80:20 & 4.16 $\pm$ 0.04 $^b$  & 4.45 $\pm$ 0.05 $^a$  & 2.23 $\pm$ 0.04 $^c$  & {<1 $\times$ 10$^{-300}$}  & {<1 $\times$ 10$^{-300}$}  & {0.998} \\
                & 70:30 & 4.52 $\pm$ 0.04 $^a$  & 4.22 $\pm$ 0.04 $^b$  & 2.45 $\pm$ 0.06 $^c$  & {<1 $\times$ 10$^{-300}$}  & {<1 $\times$ 10$^{-300}$}  & {0.997} \\
                & 60:40 & 4.40 $\pm$ 0.05 $^a$  & 4.45 $\pm$ 0.18 $^a$  & 2.78 $\pm$ 0.07 $^b$  & {2.49 $\times$ 10$^{-254}$}  & {7.47 $\times$ 10$^{-254}$}  & {0.980} \\
                & 50:50 & 4.80 $\pm$ 0.06 $^b$  & 5.20 $\pm$ 0.22 $^a$  & 3.60 $\pm$ 0.11 $^c$  & {6.87 $\times$ 10$^{-205}$}  & {2.06 $\times$ 10$^{-204}$}  & {0.958} \\
            \bottomrule
            		\end{tabularx}
            	\end{adjustwidth}

\noindent{\footnotesize{%
\textsuperscript{1} {Mean} 
 and SD were computed across 100 iterations, where each iteration involved a different randomly excluded time point from a randomly selected calf before the train--test~split. 
\textsuperscript{2} R$^{2}$ = coefficient of determination, MSE = mean squared error, RMSE = root mean square error, MAE = mean absolute error, and MAPE = mean absolute percentage~error. 
\textsuperscript{3} For each train--test split (\(90:10\), \(80:20\), \(70:30\), \(60:40\), and~\(50:50\)), the~first \(X\%\) of each calf’s time-series data was used for training, while the remaining (\(100-X\%\)) was reserved for testing. This approach ensured temporal consistency by training on earlier time points and testing on later~observations. 
{\textsuperscript{4} \textit{p}-values indicate statistical significance in model performance differences, with~values below 0.05 considered significant. }
{\textsuperscript{5} Adjusted \textit{p}-values were obtained using the Bonferroni correction to control for multiple comparisons, ensuring that statistical significance is maintained while reducing the risk of Type I errors.} 
{\textsuperscript{6} Eta squared ($\eta^2$) is an effect size measured that quantifies the proportion of variation in BW prediction performance explained by the choice of predictive model. Higher $\eta^2$ values indicate a greater influence of the model on prediction accuracy.} 
$^{a,b,c}$ superscripts compared the performance of LR, XGBoost, and~LMM models for each metric (R$^{2}$, MAE, MAPE, MSE, and RMSE). Different superscripts indicate significant differences (\textit{p} < 0.05) based on Analysis of Variance (ANOVA) followed by Tukey's Honest Significant Difference (HSD) post hoc test. For~each segmentation method (YOLOv8m and threshold), superscripts compared the performance of LR, XGBoost, and~LMM models within the same method. That is, within~YOLOv8m, LR, XGBoost, and~LMM models were compared, and~separately, within~threshold, LR, XGBoost, and~LMM models were compared.}}
\end{table}
\unskip

\begin{table}[H]
\caption{Comparison of body weight prediction performance using linear regression (LR), extreme gradient boosting (XGBoost), and~linear mixed models (LMM) with cross-validation {using the threshold segmentation}. Values are presented as mean $\pm$ standard deviation (SD){\textsuperscript{1}}. \label{tab5}}
	\begin{adjustwidth}{-\extralength}{0cm}
		\begin{tabularx}{\linewidth}{>{\centering\arraybackslash}p{0.075\linewidth} >{\centering\arraybackslash}p{0.06\linewidth} >{\centering\arraybackslash}p{0.12\linewidth} >{\centering\arraybackslash}p{0.12\linewidth} >{\centering\arraybackslash}p{0.12\linewidth} >{\centering\arraybackslash}p{0.12\linewidth} >{\centering\arraybackslash}p{0.12\linewidth} >{\centering\arraybackslash}p{0.07\linewidth}}
		  \toprule
			\textbf{Metric{\textsuperscript{2}}} & \textbf{Train:Test{\textsuperscript{3}}} & \textbf{LR} & \textbf{XGBoost} & \textbf{LMM} & \textbf{{\textit{p}-Value}{\textsuperscript{4}}} & \textbf{{Adjusted \textit{p}-Value}{\textsuperscript{5}}} & \textbf{{Eta Squared (\boldmath{$\eta^2$}){\textsuperscript{6}}}} \\
			\midrule
            \multirow{5}{*}{R$^2$} 
            & 90:10 & 0.72 $\pm$ 0.05 $^b$ & 0.73 $\pm$ 0.01 $^b$ & {0.98 $\pm$ 0.004 $^a$} & {5.06 $\times$ 10$^{-194}$} & {1.52$\times$10$^{-193}$} & {0.950} \\
            & 80:20 & 0.74 $\pm$ 0.01 $^c$ & {0.75 $\pm$ 0.002 $^c$} & {0.95 $\pm$ 0.002 $^a$} & {<1 $\times$ 10$^{-300}$} & {<1 $\times$10$^{-300}$} & {0.997} \\
            & 70:30 & 0.77 $\pm$ 0.01 $^b$ & {0.75 $\pm$ 0.004 $^c$} & {0.94 $\pm$ 0.003 $^a$} & {1.12 $\times$ 10$^{-317}$} & {3.35 $\times$ 10$^{-317}$} & {0.993} \\
            & 60:40 & 0.73 $\pm$ 0.01 $^c$ & {0.74 $\pm$ 0.002 $^c$} & {0.94 $\pm$ 0.004 $^a$} & {<1 $\times$ 10$^{-300}$} & {<1 $\times$ 10$^{-300}$} & {0.995} \\
            & 50:50 & 0.73 $\pm$ 0.01 $^c$ & {0.74 $\pm$ 0.005 $^c$} & {0.9 $\pm$ 0.005 $^a$} & {<1 $\times$ 10$^{-300}$} & {<1 $\times$ 10$^{-300}$} & {0.995} \\
            \midrule
            \multirow{5}{*}{MSE (lb$^2$)} 
            & 90:10 & 118.73 $\pm$ 6.5 $^b$ & 175.44 $\pm$ 7.21 $^a$ & 22.32 $\pm$ 1.11 $^c$ & {1.89 $\times~10^{-320}$} & {$5.67 \times 10^{-320}$} & {0.993} \\
            & 80:20 & 147.92 $\pm$ 3.32 $^a$ & 148.94 $\pm$ 1.38 $^a$ & 27.24 $\pm$ 0.87 $^b$ & {<1 $\times~10^{-300}$} & {<1 $\times~10^{-300}$} & {0.999} \\
            & 70:30 & 140.26 $\pm$ 3.62 $^b$ & 162.1 $\pm$ 2.94 $^a$ & 35.01 $\pm$ 1.12 $^c$ & {<1 $\times~10^{-300}$} & {<1 $\times~10^{-300}$} & {0.998} \\
            & 60:40 & 151.56 $\pm$ 3.3 $^b$ & 160.53 $\pm$ 1.5 $^a$ & 36.83 $\pm$ 1.29 $^c$ & {<1 $\times~10^{-300}$} & {<1 $\times~10^{-300}$} & {0.999} \\
            & 50:50 & 142.63 $\pm$ 1.49 $^b$ & 157.29 $\pm$ 2.77 $^a$ & 65.51 $\pm$ 3.63 $^c$ & {<1 $\times~10^{-300}$} & {<1 $\times~10^{-300}$} & {0.996} \\
            \midrule
            \multirow{5}{*}{RMSE (lb)} 
            & 90:10 & 10.89 $\pm$ 0.28 $^b$ & 13.24 $\pm$ 0.28 $^a$ & 4.72 $\pm$ 0.12 $^c$ & {<1 $\times~10^{-300}$} & {<1 $\times~10^{-300}$} & {0.996} \\
            & 80:20 & 12.16 $\pm$ 0.14 $^a$ & 12.20 $\pm$ 0.06 $^a$ & 5.22 $\pm$ 0.08 $^b$ & {<1 $\times~10^{-300}$} & {<1 $\times~10^{-300}$} & {0.999} \\
            & 70:30 & 11.84 $\pm$ 0.15 $^b$ & 12.73 $\pm$ 0.12 $^a$ & 5.92 $\pm$ 0.09 $^c$ & {<1 $\times~10^{-300}$} & {<1 $\times~10^{-300}$} & {0.999} \\
            & 60:40 & 12.31 $\pm$ 0.14 $^b$ & 12.67 $\pm$ 0.06 $^a$ & 6.07 $\pm$ 0.11 $^c$ & {<1 $\times~10^{-300}$} & {<1 $\times~10^{-300}$} & {0.999} \\
            & 50:50 & 11.94 $\pm$ 0.06 $^b$ & 12.54 $\pm$ 0.11 $^a$ & 8.09 $\pm$ 0.23 $^c$ & {<1 $\times~10^{-300}$} & {<1 $\times~10^{-300}$} & {0.995} \\
            \midrule
            \multirow{5}{*}{MAE (lb)} 
            & 90:10 & 8.83 $\pm$ 0.26 $^b$ & 10.56 $\pm$ 0.23 $^a$ & 3.95 $\pm$ 0.11 $^c$ & {<1 $\times~10^{-300}$} & {<1 $\times~10^{-300}$} & {0.995} \\
            & 80:20 & 9.29 $\pm$ 0.13 $^b$ & 9.88 $\pm$ 0.06 $^a$ & 4.23 $\pm$ 0.07 $^c$ & {<1 $\times~10^{-300}$} & {<1 $\times~10^{-300}$} & {0.999} \\
            & 70:30 & 9.65 $\pm$ 0.11 $^b$ & 10.06 $\pm$ 0.1 $^a$ & 4.64 $\pm$ 0.08 $^c$ & {<1 $\times~10^{-300}$} & {<1 $\times~10^{-300}$} & {0.999} \\
            & 60:40 & 9.73 $\pm$ 0.09 $^b$ & 10.22 $\pm$ 0.07 $^a$ & 4.62 $\pm$ 0.09 $^c$ & {<1 $\times~10^{-300}$} & {<1 $\times~10^{-300}$} & {0.999} \\
            & 50:50 & 9.55 $\pm$ 0.07 $^b$ & 10.19 $\pm$ 0.1 $^a$ & 6.25 $\pm$ 0.18 $^c$ & {<1 $\times~10^{-300}$} & {<1 $\times~10^{-300}$} & {0.995} \\
            \midrule
            \multirow{5}{*}{MAPE (\%)} 
            & 90:10 & 5.71 $\pm$ 0.17 $^b$ & 7.03 $\pm$ 0.17 $^a$ & 2.37 $\pm$ 0.07 $^c$ & {<1 $\times~10^{-300}$} & {<1 $\times~10^{-300}$} & {0.995} \\
            & 80:20 & 5.78 $\pm$ 0.08 $^b$ & 6.38 $\pm$ 0.05 $^a$ & 2.49 $\pm$ 0.05 $^c$ & {<1 $\times~10^{-300}$} & {<1 $\times~10^{-300}$} & {0.999} \\
            & 70:30 & 5.98 $\pm$ 0.06 $^b$ & 6.61 $\pm$ 0.07 $^a$ & 2.76 $\pm$ 0.05 $^c$ & {<1 $\times~10^{-300}$} & {<1 $\times~10^{-300}$} & {0.999} \\
            & 60:40 & 6.22 $\pm$ 0.06 $^b$ & 6.87 $\pm$ 0.06 $^a$ & 2.80 $\pm$ 0.06 $^c$ & {<1 $\times~10^{-300}$} & {<1 $\times~10^{-300}$} & {0.999} \\
            & 50:50 & 6.19 $\pm$ 0.04 $^b$ & 6.97 $\pm$ 0.09 $^a$ & 3.89 $\pm$ 0.12 $^c$ & {<1 $\times~10^{-300}$} & {$< 1 \times 10^{-300}$} & {0.996} \\

            \bottomrule
            		\end{tabularx}
            	\end{adjustwidth}

\noindent{\footnotesize{%
\textsuperscript{1} {Mean} 
 and SD were computed across 100 iterations, where each iteration involved a different randomly excluded time point from a randomly selected calf before the train--test~split. 
\textsuperscript{2} R$^{2}$ = coefficient of determination, MSE = mean squared error, RMSE = root mean square error, MAE = mean absolute error, and MAPE = mean absolute percentage~error. 
\textsuperscript{3} For each train--test split (\(90:10\), \(80:20\), \(70:30\), \(60:40\), and~\(50:50\)), the~first \(X\%\) of each calf’s time-series data was used for training, while the remaining (\(100-X\%\)) was reserved for testing. This approach ensured temporal consistency by training on earlier time points and testing on later~observations. 
{\textsuperscript{4} \textit{p}-values indicate statistical significance in model performance differences, with~values below 0.05 considered significant. }
{\textsuperscript{5} Adjusted \textit{p}-values were obtained using the Bonferroni correction to control for multiple comparisons, ensuring that statistical significance is maintained while reducing the risk of Type I errors.} 
{\textsuperscript{6} Eta squared ($\eta^2$) is an effect size measure that quantifies the proportion of variation in BW prediction performance explained by the choice of predictive model. Higher $\eta^2$ values indicate a greater influence of the model on prediction accuracy.} 
$^{a,b,c}$ superscripts compared the performance of LR, XGBoost, and~LMM models for each metric (R$^{2}$, MAE, MAPE, MSE, and RMSE). Different superscripts indicate significant differences (\textit{p} < 0.05) based on Analysis of Variance (ANOVA) followed by Tukey's Honest Significant Difference (HSD) post hoc test. For~each segmentation method (YOLOv8m and threshold), superscripts compared the performance of LR, XGBoost, and~LMM models within the same method. That is, within~YOLOv8m, LR, XGBoost, and~LMM models were compared, and~separately, within~threshold, LR, XGBoost, and~LMM models were compared.}}
\end{table}
\unskip

When using the threshold-based segmentation method, all three models exhibited lower R$^{2}$ values and higher errors, reflecting the reduced accuracy of body metric extraction compared to YOLOv8m. The~LMM method still outperformed both LR and XGBoost, with~an R$^{2}$ of 0.98 in the 90:10 split and 0.90 in the 50:50 split, maintaining a strong predictive performance despite the segmentation method. The~LR and XGBoost methods had similar predictive capabilities in the threshold-based approach, with~LR slightly outperforming XGBoost in terms of MAE and MAPE. However, both models had lower accuracy compared to their performance under YOLOv8m segmentation, with~R$^{2}$ values dropping to 0.72 (LR) and 0.73 (XGBoost) in the 90:10 split, indicating greater challenges in BW prediction when body metrics were extracted using the threshold-based approach. Notably, LMM remained stable across both segmentation methods, demonstrating its ability to model BW trends effectively even when the extracted body metrics were less reliable ({Table~\ref{tab5}}).

Overall, these findings indicate that YOLOv8m provides a more reliable segmentation method for extracting body metrics, leading to more accurate BW predictions across all models. Furthermore, LMM remains the most robust predictive model, while LR consistently outperforms XGBoost in terms of lower errors and higher R$^{2}$ values, particularly when using YOLOv8m segmentation. Additionally, the~impact of training data size was evident across all models, with~larger training sets (90:10 split) yielding better predictive performance compared to smaller ones (50:50 split). Results for other YOLO segmentation models (n, s, l, x) and BW predictions using LR, XGBoost, and~LMM are provided in Table~S2.

\section{Discussion}
This study evaluated the effectiveness of deep learning- (YOLOv8) and threshold-based image segmentation methods for predicting dairy calf BW from different time points. Our results demonstrated that YOLOv8 provided more precise body metrics, which subsequently improved BW prediction model performance. We investigated two BW prediction strategies: (1) single-time-point cross-validation using LR and XGBoost, where XGBoost consistently outperformed LR, and~(2) multiple-time-point cross-validation using LR, XGBoost, and~LMM, with~LMM proving to be the most robust and accurate~model.

The performance of BW prediction models is directly influenced by the segmentation method used to extract body metrics. Choosing between traditional threshold-based and deep learning-based segmentation methods involves a trade-off between simplicity and accuracy. Although~the threshold approach is computationally efficient, it struggled with complex backgrounds, overexposure, and underexposure due to the camera's sensitivity to lighting. In~this study, the~threshold method segmented only 573 of 957 images due to overexposure and/or the proximity of the calf to the automatic feeder boundaries. Determining the optimal threshold value was challenging—lower thresholds risked including external boundaries, while higher thresholds excluded parts of the calf’s body, leading to incomplete contours. Moreover, manual video collection often resulted in low-quality data. This is because video collection time was decided by human availability rather than natural feeding times, leading to varied calf postures like raised heads. Ideally, calves should be recorded in a consistent drinking posture, underscoring the need for an automated system to capture images during feeding, which could improve both image quality and animal welfare. However, even with automation, occasional low-quality images from cameras can still occur under ideal lighting conditions. In~such cases, deep learning models, like YOLOv8, effectively adjust for such variability, as~demonstrated in our study. The~model YOLOv8 was trained on diverse datasets with different lighting conditions and low-quality images, enabling robust segmentation, even in suboptimal scenarios. Unlike the threshold method, YOLOv8 exhibited greater resilience to imaging challenges and consistently provided more accurate segmentations. However, YOLOv8 requires larger datasets and advanced computational resources, making it more resource-intensive. Despite these differences, our findings suggest that integrating automated image collection with deep learning segmentation enhances segmentation accuracy and ultimately improves BW prediction~performance.

{Previous research has explored imaging techniques and machine learning models for calf BW prediction. \mbox{\citet{song2014body}} used a dorsal-view PMDTM 3D ToF camera angled 10° downward to capture Holstein calves at an automatic feeder. This setup required angle adjustments to convert 3D data into real-world coordinates, adding complexity. By contrast, our horizontal camera setup avoided this step, reducing errors and simplifying body metric measurements. \mbox{\citet{song2014body}} used 3D depth data to segment calf bodies by removing background noise and applying height thresholds. Their study included 49 Holstein calves for training and 19 for validation, extracting features like body volume, hip height, and height distribution for BW prediction. In contrast, our study used OpenCV for threshold-based segmentation on the hue channel of HSV color space, with dynamic contour detection based on shape matching and contour area. \mbox{\citet{song2014body}}'s model, which used multiple LR, achieved a mean relative error of 6.50\% and an RMSE of 13.67 lb (6.20 kg) on validation data. \mbox{\citet{nishide2018calf}} collected lateral-view images of Japanese Black calves using two 3D cameras, estimating body area through background subtraction and a 3D cylindrical model. Their model achieved an R$^{2}$ of 0.91 and a MAPE of 6.39\%, but the results are limited by the lack of validation on an unseen dataset. Additionally, all calves in \mbox{\citet{nishide2018calf}} study were black, ideal for data collection, whereas our Holstein calves had black and white patterns, with white areas prone to overexposure, affecting segmentation. Similarly, a recent study used the YOLOv8 network to predict the live BW of yak heifers by detecting normal postures and extracting body size parameters from dorsal-view 2D images \mbox{\cite{peng2024dynamic}}. Their model achieved a high R$^{2}$ of 0.96 and an RMSE of 2.43 kg, but images of the same 16 yaks were likely included in both training and testing, potentially inflating the model’s performance. This highlights the importance of ensuring non-overlapping animals in training and testing to avoid overfitting. Given that real-world applications require models to predict BW for unseen calves rather than memorizing specific individuals, maintaining a clear separation between training and testing datasets is essential. Cross-validation is a widely adopted technique to enhance model robustness and prevent data leakage by systematically partitioning the dataset. In our study, a 5-fold cross-validation strategy was applied to LR and XGBoost models, ensuring that training and testing sets consisted of different calves, thereby strengthening model generalization.}

While the single-time-point cross-validation provides a robust framework for evaluating model generalization across different calves, it does not account for temporal dependencies in BW changes over time. To~address this, a~longitudinal analysis was conducted to model individual growth trajectories and forecast BW at future time points. Our study demonstrated strong correlations between body metrics and BW across different age groups. This suggests that body size measurements obtained at younger ages can serve as reliable predictors for BW at later stages. One study found that birth weight and early milk intake influenced BW at 400 days, with~each kilogram at birth adding 2.5 kg at 400 days. Calves with moderate milk intake in the first 60 days had higher predicted weights~\cite{hurst2021predictive}. Additionally, a~study developed a statistical model to predict live weight in young male dairy calves aged between 10 and 42 days, incorporating factors like dam parity, gestation length, and~parental average genetic merit. This model achieved a correlation of 0.76 between actual and predicted live weights, suggesting that genetic and early-life factors can be utilized to estimate future growth trajectories~\cite{dunne2021predicting}. {Building on these findings, our study further validated the predictive capability of body metrics measured at early time points through longitudinal BW forecasting using LR, XGBoost, and~LMM.} {The LMM method consistently outperformed other models across all train--test splits, likely because it accounts for time-dependent variance by modeling both fixed and random effects. In~contrast, XGBoost assumes that each BW measurement is independent, leading to overfitting short-term fluctuations rather than capturing long-term growth trends. This limitation became more evident in lower train--test splits, where LMM maintained stability while XGBoost struggled with reduced training data. Moreover, LR also outperformed XGBoost, suggesting that BW follows a relatively structured trend where a linear relationship with body metrics provides strong predictive power. Unlike XGBoost, which attempts to capture complex nonlinear patterns, LR maintains a stable, global trend, avoiding unnecessary noise. These results indicate that models preserving temporal dependencies perform better in BW prediction than those relying solely on decision trees.} Notably, a~study on dairy cows demonstrated that integrating the Mask Region-Based Convolutional Neural Network (Mask R-CNN) approach with LMM achieved the best prediction performance, with~an R$^2$ of 0.98 and a MAPE of 2.03\% using a 90:10 split~\cite{bi2023depth}. Similarly, our study found that combining deep learning-based segmentation (YOLOv8) with LMM for BW prediction in a 90:10 split achieved the best performance, with~an R$^2$ of 0.99 and a MAPE of 2.39\%. {These findings highlight the effectiveness of integrating advanced modeling approaches, such as deep learning-based segmentation and LMM, for~accurate BW prediction, reinforcing the potential of body metrics measured at early time points in forecasting long-term growth trajectories.}

Limited research has focused on image-based BW prediction for Holstein calves, with~most existing studies relying on morphometric measurements~\cite{ozkaya2012accuracy} or focusing on beef breeds such as Angus-cross~\cite{rotondo2021predicting}, yak~\cite{peng2024dynamic}, and~Nellore~\cite{cominotte2020automated}. Additionally, most studies on Holsteins have centered on adult cows rather than pre-weaned calves~\cite{martins2020estimating, bi2023depth}. {This gap in research highlights the importance of considering breed-specific factors, as~differences in coat patterns and body structures influence both segmentation and BW prediction accuracy. For~instance, Holstein calves present unique challenges due to their black-and-white coat patterns, which increase the likelihood of overexposure or underexposure in images, making segmentation more complex. Our approach, which integrates dorsal depth images with YOLOv8-based segmentation, effectively addresses these challenges and improves BW prediction accuracy under real farm conditions. Another breed-specific factor that influences both segmentation and BW prediction is that of structural differences, such as Brahman humps or Belgian Blue’s double muscling, which further affect contour extraction. These anatomical variations also impact BW prediction, as~dairy breeds like Holstein and Jersey have leaner frames, while beef breeds such as Charolais and Limousin exhibit greater muscle mass, altering the relationship between body contour and actual weight.} {Furthermore, limited studies have explored non-contact measurements taken at early time points for predicting later BW, despite their significance for calf health and welfare.} The early identification of growth deviations enables timely intervention, optimizing feeding strategies and reducing metabolic disorder risks. Traditional methods relying on direct weighing or manual morphometric assessments are labor-intensive and can induce stress. By~leveraging automated image-based BW tracking, our study provides a scalable, non-invasive alternative that enhances precision livestock management and minimizes animal handling~stress.

The primary application of this study is to advance automated BW monitoring in dairy calves, supporting precision livestock farming. By~integrating deep learning-based segmentation (YOLOv8) with BW prediction models (LMM and XGBoost), this approach enables real-time, non-invasive weight estimation, reducing the need for manual handling. Automated BW tracking facilitates early intervention for calves deviating from expected growth trajectories, improving nutritional management and herd health. Beyond~dairy calves, this methodology has potential applications in other livestock species, including pigs~\cite{pezzuolo2018barn}, sheep~\cite{liang2024study}, and~goats~\cite{li2022study}, enabling broader advancements in automated growth monitoring and livestock~management.

Future research should aim to validate the developed models across diverse breeds, age groups, and~environmental conditions to enhance their robustness and applicability. Exploring the integration of additional data sources, such as genetic information and health records, could further improve BW prediction accuracy. {Given the computational demands of YOLOv8, future studies should explore techniques such as model pruning, quantization, and~lightweight architectures to enhance efficiency and facilitate broader adoption, particularly for on-farm applications with limited computational resources.} Additionally, investigating the use of alternative imaging technologies, such as thermal imaging or 2D cameras, may offer new insights and improve segmentation accuracy under varying farm~conditions.

\section{Conclusions}

In conclusion, this study demonstrated that deep learning-based segmentation, specifically YOLOv8, significantly improves the accuracy of BW prediction in dairy calves compared to traditional threshold-based methods. By~integrating single-time-point cross-validation and multiple-time-point cross-validation, we established a robust predictive framework. In~single-time-point cross-validation, XGBoost consistently outperformed LR, making it a valuable tool for BW prediction when individual growth trends are not a primary concern. However, for~multiple-time-point cross-validation, LMM proved to be the most reliable model, effectively capturing individual growth trajectories over time. {The strong correlations between body metrics measured at early time points and later BW further emphasize their potential as effective indicators for future growth, aiding in better management decisions for dairy operations. These advancements in BW prediction contribute to the development of precision livestock farming, offering data-driven solutions to enhance animal monitoring, optimize resource allocation, and~improve overall herd health and {productivity.} 
}

\vspace{6pt} 

\supplementary{{~~~~~~~~~~~~~~~} 
}



\authorcontributions{Conceptualization, R.R.C. and M.L.; methodology, M.L. G.M., Y.B., and R.R.C.; validation, M.L.; investigation, M.L.; resources, M.L., G.M., and~Y.B.; data curation, M.L.; writing---original draft preparation, M.L.; writing---review and editing, R.R.C., G.M., and~Y.B.; visualization, M.L.; supervision, R.R.C.; project administration, M.L.; funding acquisition, R.R.C. All authors have read and agreed to the published version of the~manuscript.}

\funding{This research was supported by John Lee Pratt Animal Nutrition Program (23383) and Virginia Agricultural Council (23185).}

\institutionalreview{The animal procedures were approved by the Virginia Polytechnic Institute and State University Animal Care and Use Committee (Protocol \#22-197).}

\informedconsent{Not applicable}

\dataavailability{The data presented in this study are available on request from the corresponding author due to legal restrictions.} 




\acknowledgments{The authors appreciate Allison Tezak, Kaylee Farmer, Kayla Alward, and~John McGehee for their help with animal~handling.}

\conflictsofinterest{The authors declare no conflicts of~interest.}

\begin{adjustwidth}{-\extralength}{0cm}

\reftitle{References}


\bibliographystyle{mdpi}

\PublishersNote{}
\end{adjustwidth}
\end{document}